\documentclass[fleqn,11pt]{article}
\usepackage[cp1251]{inputenc}
\usepackage[english]{babel}
\usepackage{amsmath,amssymb}
\usepackage{xcolor}
\usepackage{verbatim}
\usepackage{dsfont}
\usepackage{graphicx}
\newtheorem{Mark}{Remarks}
\newtheorem{Conclusion}{Conclusions}

\newcommand{\sgn}{\mathrm{sgn}}

\newcommand{\Fig}[3]{%
\begin{center}
\parbox{#2cm}{%
\includegraphics[width=#2cm,width=7cm]{#1}}\\
\parbox{8.5cm}{%
\noindent {\refstepcounter{figure} \small \textbf{Fig. \thefigure.}\;
#3}}\end{center}}
\newcommand{\FigH}[4]{%
\begin{flushleft}
\parbox{#2cm}{%
\includegraphics[width=#2cm,height=#3cm]{#1}}
\parbox{8.5cm}{%
\noindent \refstepcounter{figure}\small \textbf{Fig. \thefigure.}\;
#4}\end{flushleft}}
\newcounter{strochka}
\newcommand{\stroka}[1]{\refstepcounter{strochka}\par\noindent\textsl{\Roman{spisok}.\arabic{strochka}}. \; \textsl{#1}}
\newcounter{spisok}
\newcommand{\spisok}[2]{%
\vspace{10pt}
\setcounter{strochka}{0}
\noindent \textbf{Remarks \refstepcounter{spisok} \Roman{spisok} #1.}

#2\hfill\vspace{10pt}
}
\newcommand{\String}[1]{\noindent $\bullet$ \hskip 6pt {\sl #1}}
\usepackage{amsfonts,amssymb,cite}
\usepackage{graphicx}

\def\noi{\noindent}

\newcommand{\Title}[1]{\noi {{\Large\bf #1}}\\[1ex]}

\newcommand{\Author}[2]{\noi{\bf #1}\\[2ex]\noi{\normalsize\it #2}\\}

\newcommand{\Abstract}[1]{\vskip 2mm \begin{center}
        \parbox{16.4cm}{\small\noi #1} \end{center}\medskip}
\newcommand{\foom}[1]{\protect\footnotemark[#1]}

\def\nqq{\hspace*{-2em}}

\usepackage{color}

\def\Jl#1#2{#1 {\bf #2},\ }

\def\ApJ#1 {\Jl{Astroph. J.}{#1}}
\def\CQG#1 {\Jl{Class. Quantum Grav.}{#1}}
\def\DAN#1 {\Jl{Dokl. AN SSSR}{#1}}
\def\GC#1 {\Jl{Grav. Cosmol.}{#1}}
\def\GRG#1 {\Jl{Gen. Rel. Grav.}{#1}}
\def\IJMPD#1 {\Jl{Int. J. Mod. Phys. D}{#1}}
\def\JETF#1 {\Jl{Zh. Eksp. Teor. Fiz.}{#1}}
\def\JETP#1 {\Jl{Sov. Phys. JETP}{#1}}
\def\JHEP#1 {\Jl{JHEP}{#1}}
\def\JMP#1 {\Jl{J. Math. Phys.}{#1}}
\def\NPB#1 {\Jl{Nucl. Phys. B}{#1}}
\def\NP#1 {\Jl{Nucl. Phys.}{#1}}
\def\PLA#1 {\Jl{Phys. Lett. A}{#1}}
\def\PLB#1 {\Jl{Phys. Lett. B}{#1}}
\def\PRD#1 {\Jl{Phys. Rev. D}{#1}}
\def\PRL#1 {\Jl{Phys. Rev. Lett.}{#1}}

\def\lal{&&\nqq {}}

\def\beq{\begin{equation}}
\def\eeq{\end{equation}}
\def\bear{\begin{eqnarray}}
\def\bearr{\begin{eqnarray} \lal}
\def\ear{\end{eqnarray}}
\def\earn{\nonumber \end{eqnarray}}

\begin{document}
\thispagestyle{empty}
\twocolumn[
%\jnumber{4}{2019}

\vspace{1cm}

\Title{Cosmological models based on an asymmetric scalar doublet with kinetic coupling of components. II. Numerical modeling\foom 1$^, $\foom 2}

\Author{Yu.G. Ignat'ev$^1$}
    {Institute of Physics, Kazan Federal University, Kremlyovskaya str., 16A, Kazan, 420008, Russia}

 \Author{I.A. Kokh$^2$}
    {N.I. Lobachevsky Institute of Mathematics and Mechanics, Kazan Federal University, Kremlyovskaya str., 35, Kazan, 420008, Russia}

\Abstract
 {Numerical modeling of a mathematical model of the cosmological evolution of an asymmetric scalar doublet with kinetic interaction between the components was carried out. A wide range of values of fundamental parameters and initial conditions of the model are considered. Various types of behavior have been identified: models with an infinite inflationary past and future - with and without a rebound point, models with a finite past and infinite future, with an infinite past and finite future (Big Rip), as well as models with a finite past and future. Based on numerical analysis, the behavior of models near the initial singularity and the Big Rip is studied; it is shown that in both cases the barotropic coefficient tends to unity, which corresponds to an extremely rigid state of matter near singularities. A numerical example of the cosmological generation of the classical component of a scalar doublet by its phantom component is given. An assessment was made of the creation of the velocity of fermion pairs by a scalar field near the rebound points and it was shown that a scalar field at the cold stage of the Universe can ensure the creation of the required number of massive scalarly charged fermions.
 \\[8pt]
}
\bigskip

] %%%%%%%%%%%%%%%%%%%%%%%%%%%

\section{Introduction}
In the first part of the article \cite{Part1} a cosmological model based on an asymmetric scalar doublet with kinetic coupling of components was formulated and its main properties were investigated. In addition, this work demonstrated an example of numerical modeling for a special case of a set of fundamental constants and initial conditions, illustrating the analytical properties of the model.

In this part of the article we will present a wider range of numerical modeling results and their analysis in order to identify types of behavior of the model, as well as solve the problem of generating one of the components of a scalar doublet by another, similar to cosmological models with scalarly charged fermions \cite{TMF_21}, \cite{ Ignat_Sasha_Dima1}, \cite{Ignat_Sasha_Dima2}. As we noted in \cite{Part1}, we will also pursue the goal of possibly replacing the mathematically complicated and cumbersome model of scalarly charged fermions with a simpler mathematical model in the theory of scalar-gravitational instability.

We present the necessary information from the first part of the article \cite{Part1}\footnote{In the future, when referring to the results of the first part of the article, we will append the Roman numeral I to the link, for example, Fig.I.15.}.

The dynamical system corresponding to the cosmological model in the spatially flat Friedmann metric with the scale factor $a(t)$ consists of a normal system of ordinary differential equations:
\begin{eqnarray}
\label{sys0a}
\dot{\xi}= H \qquad (a\equiv \exp(\xi));\\
\label{sys1a}
\dot{\Phi}=  Z; \\
\label{sys2a}
\dot{Z}  =  -3HZ-\frac{m^2\Phi -\alpha \Phi ^{3}}{1+\gamma^2}-\gamma\frac{\mu^{2}\varphi -\beta \varphi ^{3}}{1+\gamma^2};\\
\label{sys3a}
\dot{\varphi}=z;\\
\label{sys4a}
\dot{z} = -3Hz+\frac{\mu^{2}\varphi -\beta \varphi ^{3}}{1+\gamma^2}- \gamma \frac{m^2\Phi -\alpha \Phi ^{3}}{1+\gamma^2};\\
\label{sys5a}
\dot{H} = -\frac{1}{2}Z^2+\frac{1}{2}z^2-\gamma Zz,
\end{eqnarray}
and its first \emph{zero} integral - the equations of the Einstein - Higgs hypersurface\footnote{In fact, the Einstein equations are $^4_4$.} of the autonomous subsystem $S_1=$ \{\eqref{sys1a} -- \eqref{sys5a}\} dynamic system $S_0=$ \{\eqref{sys0a} -- \eqref{sys5a}\} ($S_1\subset S_0$):
\begin{equation}\label{Hip_Ein}
\begin{array}{l}
3H^2-\dfrac{Z^2}{2}+\dfrac{\alpha\Phi^4}{4}-\dfrac{m^2\Phi^2}{2}\\[11pt]
+\dfrac{z^2}{2}+\dfrac{\beta\varphi^4}{4}-\dfrac{\mu^2\varphi^2}{2}
-\gamma Zz-\Lambda=0,
\end{array}
\end{equation}
where $\Phi(t)$ and $\varphi(t)$ are the potentials of classical and phantom scalar fields, $H(t)$ is the Hubble parameter, $\alpha,\beta$ are the self-interaction constants of these fields, $m,\mu$ are the masses of their quanta, $\gamma$ is the kinetic interaction constant of the classical and phantom scalar fields.
\begin{equation}\label{L->L0}
\Lambda=\Lambda_0-\frac{m^4}{4\alpha}-\frac{\mu^4}{4\beta},
\end{equation}
and $\Lambda_0$ is the seed value of the cosmological constant.

Since both dynamical systems, $S_0$ and $S_1$, are autonomous, in cases where we do not need knowledge of the scale factor function $a(t)$, we will study the autonomous subsystem $S_1$, otherwise - the system $S_0$. Let us note in passing that, in principle, in all cases it would be possible to limit ourselves to studying the system $S_1$, since the scale factor $a(t)$ using the equation \eqref{sys0a} could be found using integration
\[a(t)=\mathrm{e}^{\int H(t)dt},\]
however, from a technical point of view, it is easier to find a numerical solution to a normal system of 6 differential equations $S_0$ than to find the integral of the function $H(t)$ based on a numerical solution to a normal system of 5 differential equations $S_1$ using numerical methods.

Further, the invariant cosmological acceleration $\Omega(t)$ is related to the Hubble parameter $H(t)$ by the relation
\begin{equation}\label{Omega}
\Omega=\frac{a\ddot{a}}{{\dot{a}}^2}\equiv 1+\frac{\dot{H}}{H^2}=-\frac{1}{2}(1+3\kappa),
\end{equation}
where $\kappa(t)$ is the barotropic coefficient of cosmological matter:
\begin{equation}\label{kappa}
\kappa=P\diagup E\equiv -\frac{1}{3}(1+2\Omega),
\end{equation}
and $E$ is \emph{normalized effective energy} and $P$ is \emph{normalized effective pressure}\footnote{multiplied by $8\pi$}:
\begin{eqnarray}\label{E_m}
E=&\displaystyle\frac{Z^2}{2}-\frac{\alpha\Phi^4}{4}+\frac{m^2\Phi^2}{2}\hskip 2cm\nonumber\\
&\displaystyle-\frac{z^2}{2}-\frac{\beta\varphi^4}{4}+\frac{\mu^2\varphi^2}{2}
+\gamma Zz+\Lambda,\\
\label{P_m}
P=&\displaystyle\frac{Z^2}{2}+\frac{\alpha\Phi^4}{4}-\frac{m^2\Phi^2}{2}\hskip 2cm\nonumber\\
&\displaystyle-\frac{z^2}{2}+\frac{\beta\varphi^4}{4}-\frac{\mu^2\varphi^2}{2}
+\gamma Zz-\Lambda,
\end{eqnarray}
at that
\begin{equation}
E+P=Z^2-z^2+2\gamma Zz\equiv-2\dot{H}.
\end{equation}
Along with the total effective energy and pressure, we introduce similar quantities for individual components of the cosmological system, normalized energy densities $E_c,E_f$ and pressure $P_c,P_f$
\begin{eqnarray}
E_c=\frac{Z^2}{2}-\frac{\alpha\Phi^4}{4}+\frac{m^2\Phi^2}{2};\\
E_f=-\frac{z^2}{2}-\frac{\beta\varphi^4}{4}+\frac{\mu^2\varphi^2}{2};\\
P_c=\frac{Z^2}{2}+\frac{\alpha\Phi^4}{4}-\frac{m^2\Phi^2}{2};\\
P_f=-\frac{z^2}{2}+\frac{\beta\varphi^4}{4}-\frac{\mu^2\varphi^2}{2}
\end{eqnarray}
and the corresponding interaction values of their components:
\begin{eqnarray}
E_{cf}=P_{cf}=\gamma Zz,
\end{eqnarray}
so that
\begin{eqnarray}
E=E_c+E_f+E_{cf}+\Lambda;\\
P=P_c+P_f+P_{cf}-\Lambda.
\end{eqnarray}
A particular cosmological model $\textbf{M}$ is determined,
firstly, by a system of dynamic equations \eqref{sys0a}--\eqref{sys5a},
secondly, by fundamental parameters
\begin{equation}\label{P}
\textbf{P} = [[\alpha,\beta, m, \mu],\gamma,\Lambda]
\end{equation}
and, thirdly, by initial conditions
\begin{equation}\label{I}
\textbf{I} = [ \Phi_0,Z_0,\varphi_0, z_0,e],\qquad (e=\pm 1).
\end{equation}

Here the sign $e$ corresponds to the choice of a positive or negative solution to the equation \eqref{Hip_Ein} relative to the initial value of the Hubble parameter. Note that using the autonomy of the dynamical systems $S_0$ and $S_1$ and the invariance of the Friedman metric with respect to time translations $t\to t+t_0$, we can always choose the initial condition for the scale function $\xi(t)$ in the form ( see for example \cite{TMF20})
\begin{equation}\label{xi(0)}
\xi(0)=0.
\end{equation}
The cosmological model under study can be considered as a five-dimensional dynamical system in the arithmetic phase space $\mathds{R}^5=\{\Phi,Z,\varphi,z,H\}=\mathds{R}^3 \cup \mathds{ R}^3$, three-dimensional phase subspaces $\mathds{R}^3=\{\Phi,Z,H\}$ and $\mathds{R}^3=\{\varphi,z,H\}$ We will further denote by the symbols $\Sigma_{\Phi}$ and $\Sigma_{\varphi}$ and, for simplicity, call them the classical and phantom phase spaces, respectively. In this case, $\Sigma_{\Phi}\cup\Sigma_{\varphi}=\mathds{R}^5$ and $\Sigma_{\Phi}\cap\Sigma_{\varphi}=\mathds{R}^1 =OH$.

The singular points of the dynamical system under study have coordinates (see\cite{Part1}):
\begin{eqnarray}\label{M00}
M_{0,0}^{\pm}\left(0,0,0,0,\pm\frac{\sqrt{3\Lambda}}{3}\right);\\
\label{M01}
M_{0,\pm 1}^{\pm} :\left(0,0,\pm \frac{\mu}{\sqrt{\beta}},0,
\pm\frac{\sqrt{3\Lambda_{\beta}}}{3}\right);\\
\label{M10}
M_{\pm 1,0}^{\pm} :\left(\pm \frac{m}{\sqrt{\alpha}},0,0,0,
\pm\frac{\sqrt{3\Lambda_{\alpha}}}{3}\right);\\
\label{M11}
M_{\pm 1,\pm 1}^{\pm} :\left(\pm \frac{m}{\sqrt{\alpha}},0,\pm \frac{\mu}{\sqrt{\beta}},0,
\pm\frac{\sqrt{3\Lambda_0}}{3}\right),
\end{eqnarray}
where
$\Lambda_{\beta}\equiv\Lambda+\mu^4/4\beta$, $\Lambda_{\alpha}\equiv\Lambda+m^4/4\alpha$.

Before moving on to the results of numerical modeling, we make the following remarks:

\spisok{(preliminary)}{\label{Mark1}%
\stroka{\label{mark1}
According to Property I.1 of the invariance of a dynamical system with respect to changes in the sign of $\gamma$, it is sufficient to study mathematical models with a positive value of $\gamma\geqslant0$;}
\stroka{\label{mark2} According to Property I.2 when transforming similarity
:
\begin{eqnarray}\label{sim1}
{\bf\tilde{P}} = \left[\left[\frac{\alpha}{k^2},\frac{\beta}{k^2}, \frac{m}{k}, \frac{\mu}{k}\right],\gamma,\frac{\Lambda}{k^2}\right]
\end{eqnarray}
solution of the Cauchy problem with initial conditions
\begin{eqnarray}\label{tilde_I}
{\bf \tilde{I}} = \left[\Phi_0,\frac{Z_0}{k},\varphi_0 ,\frac{z_0}{k},e\right]
\end{eqnarray}
is obtained from the solution $\mathbf{S}$ of the Cauchy problem for the preimage according to the rule
\begin{eqnarray}\label{sim3}
\textbf{S}(t)= [\Phi(t), Z(t), \varphi(t), z(t), H(t)] \Rightarrow\\
\!\!\!\mathbf{\tilde{S}}(t)=\left[ \Phi\left(t\right),\frac{1}{k}Z\left(t\right),\varphi\left(t\right),\frac{1}{k}z\left(t\right),\frac{1}{k} H\left(t\right)\right],
\end{eqnarray}
where the connection between the time $\tilde{t}$ in the model ${\bf\tilde{M}}$ and the time $t$ of the preimage ${\bf M}$ is taken into account
\begin{equation}\label{sim2}
\tilde{t}=kt;
\end{equation}
\vspace{-12pt}
}
\stroka{\label{mark3} According to Property I.4, the coordinates of singular points of an autonomous dynamic system \eqref{sys1a}--\eqref{sys5a} with the integral condition \eqref{Hip_Ein}, as well as their character, coincide with the coordinates of singular points
and their character for a dynamic system without interaction of components \cite{TMF20};
}
\stroka{\label{mark4}According to \cite{TMF_24} during the similarity transformation \eqref{sim1} -- \eqref{sim2} the eigenvalues $\lambda$ of the matrix of the dynamical system are transformed according to the rule
\begin{equation}\label{tilde_lambda}
\tilde{\lambda}=\frac{\lambda}{k};
\end{equation}
}
\stroka{\label{mark5}All singular points \eqref{M00} -- \eqref{M11} of the dynamical system $S_0$ are exact constant solutions of this system. Wherein
\begin{equation}\label{a=exp}
a(t)=e^{\pm H_0 t},
\end{equation}
where $H_0$ is the positive value of the 5th coordinate of the corresponding singular point from the list \eqref{M00} -- \eqref{M11}.
Thus, all singular points correspond to inflationary solutions - inflationary expansion when choosing a positive sign in \eqref{a=exp}, or inflationary compression when choosing a negative sign. In both cases, these solutions correspond to cosmological models with an infinite past and an infinite future.
At the same time, the question of the sustainability of these solutions remains open.}}
\section{Basic model}
Let's set the parameters of the basic model as follows:
\begin{equation}\label{P_0}
\mathbf{P_0}=[[1, 1,1,1],\gamma,3\cdot10^{-6}],
\end{equation}
where the constant $\gamma$ is still an arbitrary parameter.
\subsection{Singular points}
The coordinates of the singular points of the model \eqref{P_0} and their characters are indicated in Tab.\ref{Tab1}. Here and in what follows, the following notations are used: $\mathbf{A}$ -- attraction, $\mathbf{R}$ -- repulsion, $\mathbf{S}$ -- saddle, we also somewhat simplify the characteristics of singular points, preserving only their properties of attraction and repulsion.\\[12pt]
\spisok{(about the character of the points)}{\label{Mark2}%
{\sl The value of the parameter $\gamma$ does not affect either the coordinates of singular points or their character.
In this case, however, the absolute values of the real and imaginary parts of the eigenvalues depend on the value of $\gamma$, but the signs of these parts do not change}.
}
\begin{center}
\refstepcounter{table} {\large Tab. \thetable. \label{Tab1} Characters (type) of singular points\linebreak in subspaces $[\Sigma_\Phi,\Sigma_\varphi]$} in the model with parameters $\mathbf{P_0}$\\[12pt]
\begin{tabular}{|c|c|c|}
\hline
Singular points & Coordinates & Type\\
\hline
$M_{0,0}^{+}$ & $[0,0,0,0,0.001]$ &  $[\mathbf{A},\ \mathbf{S}]$ \\
\hline
$M_{0,0}^{-}$ & $[0,0,0,0,-0.001]$ & $[\mathbf{R},\ \mathbf{S}]$\\
\hline
$M_{0,\pm1}^{+}$ & $[0,0,\pm1,0,0.289]$ & $[\mathbf{A},\ \mathbf{A}]$ \\
\hline
$M_{0,\pm1}^{-}$ &  $[0,0,\pm1,0,-0.289]$ & $[\mathbf{R},\ \mathbf{R}]$ \\
\hline
$M_{\pm1,0}^{+}$ & $[\pm1,0,0,0,0.289]$ & $[\mathbf{S},\ \mathbf{S}]$ \\
\hline
$M_{\pm1,0}^{-}$ & $[\pm1,0,0,0,-0.289]$ &   $[\mathbf{S},\ \mathbf{S}]$ \\
\hline
$M_{\pm1,\pm1}^{+}$ & $[\pm1,0,\pm1,0,0.409]$ & $[\mathbf{S},\ \mathbf{A}]$ \\
\hline
$M_{\pm1,\pm1}^{-}$ & $[\pm1,0,\pm1,0,-0.409]$ &  $[\mathbf{S},\ \mathbf{R}]$ \\
\hline
\end{tabular}
\end{center}
Due to Remarks II, the results of Tab. \ref{Tab1} are applicable to models with arbitrary values of the $\gamma$.

\subsection{System in steady state}
Let us now specify the $\gamma$ parameter and consider a model with the following parameters:
\begin{equation}\label{P_1}
\mathbf{P_1}=\bigl[[1, 1,1,1],10^{-5},3\cdot10^{-6}\bigl].
\end{equation}
According to Tab.\ref{Tab1}, there are only two singular points, $M_{0,\pm1}^{+}$ (see \eqref{M01}), which are simultaneously attractive in the subspaces $\Sigma_{\Phi} $ and $\Sigma_{\varphi}$.

In Fig.\ref{Ign2_1} shows the evolution of geometric factors - the scale function $\xi(t)$ and the Hubble parameter $H(t)$ for a model with parameters $\mathbf{P_1}$ under initial conditions corresponding to the coordinates of stable points $M_{0,\pm1}^{+}$
\begin{equation}\label{I_0}
\mathbf{I^\pm_0}=[0,0,0,1,\pm1].
\end{equation}
\FigH{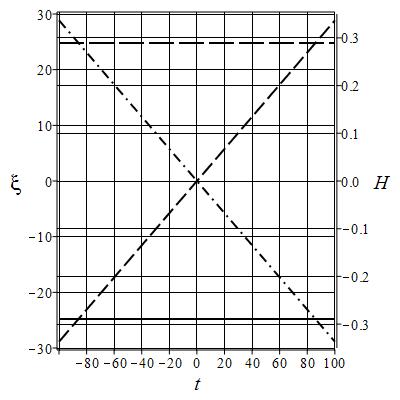}{8}{6.5}{\label{Ign2_1}Functions $\xi(t)$ are dashed and dashed-dotted, $H(t)$ are solid and long-dashed lines, respectively, for the initial conditions $\mathbf{I^+_0},\mathbf{I^-_0}$.}
In this case, the initial conditions coincide with the coordinates of the stable singular point, the exact solution is the constant scalar potentials $\Phi_0=0,\,\varphi_0=\pm \mu/\sqrt{\beta}$. In this case, the model corresponds to either an infinite inflationary expansion ($H=H_+>0$) or an infinite inflationary contraction ($H=H_-<0$).

\subsection{Rebound case}
If the initial conditions do not coincide with the coordinates of stable singular  points, a rebound point appears in the models -- the minimum of the scale function $\xi(t)\Rightarrow H=0$.
At this point, inflationary compression gives way to inflationary expansion with a symmetrical value of the Hubble constant parameter $H_\pm=\pm H_0$.
\FigH{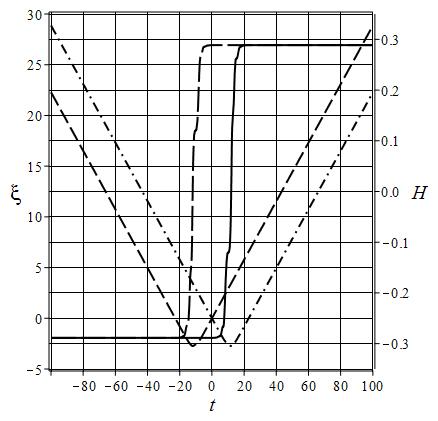}{8}{6.5}{\label{Ign2_2}Functions $\xi(t)$ are dashed and dash - dotted, $H(t)$ are solid and long-dashed lines, respectively for initial conditions $\mathbf{I^+_1}$ and $\mathbf{I^-_1}$.}
\FigH{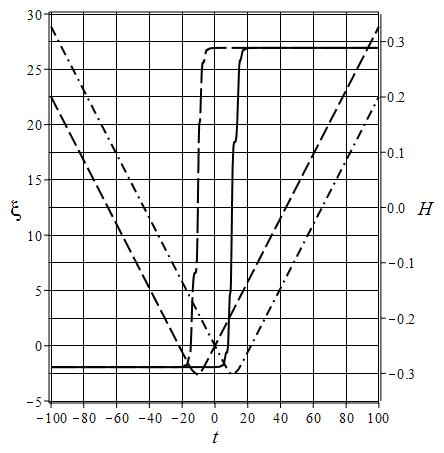}{8}{6.5}{\label{Ign2_3}Functions $\xi(t)$ are dashed and dash - dotted, $H(t)$ are solid and long-dashed lines, respectively for initial conditions $\mathbf{I^+_2}$ and $\mathbf{I^-_2}$.}
In Fig.\ref{Ign2_2}--\ref{Ign2_3} shows the evolution of geometric factors - the scale function $\xi(t)$ and the Hubble parameter $H(t)$ for a model with parameters $\mathbf{P_1}$ under initial conditions, close to the coordinates of stable points $M_{0,\pm1}^{+}$
\begin{equation}\label{I_+}
\begin{array}{l}
\mathbf{I^\pm_1}=[0,0,0,0.99,\pm1],\\
\mathbf{I^\pm_2}=[0,0,0,1.01,\pm1].
\end{array}
\end{equation}

Commenting on the graphs in Fig.\ref{Ign2_1}--\ref{Ign2_3}, note the following:

\String{All the cases considered above describe cosmological models with an infinite past and an infinite future.}\\[3pt]
\String{In this case, however, in the case of the initial conditions \eqref{I_0} coinciding with the coordinates of the stable singular point, the entire cosmological history is described by the inflationary solution corresponding
limitless expansion or contraction of the Universe. In this case, the graphs of the functions $\xi(t)$ and $H(t)$ do not depend on the value of the parameter $\gamma$.}\\[3pt]
\String{At the slightest discrepancy between the initial conditions and the coordinates of the stable singular point $M_{0,\pm1}^{+}$, the cosmological history breaks down, regardless of the initially given sign of the Hubble parameter
into two stages: first - inflationary compression, then - inflationary expansion. At the same time, graphs of the geometric factors $\xi(t)$ and $H(t)$ corresponding to the initial conditions with positive and negative
sign of the Hubble parameter, are similar with a shift along the time axis.}\\[12pt]
\subsection{Evolution of scalar fields}
If the initial conditions coincide with the coordinates of the stable singular point, the exact solution, as we indicated above, is the constant scalar potentials $\Phi_0=0,\varphi_0=\pm \mu/\sqrt{\beta}$. Therefore, we can talk about the evolution of scalar fields only when the initial state deviates from a stable singular point.

In Fig.\ref{Ign2_6}--\ref{Ign2_7} shows graphs of the evolution of scalar potentials $\Phi(t)$ and $\varphi(t)$ for initial conditions $\mathbf{I^+_1}$ and $\mathbf{ I^+_2}$. From these graphs it is clear that the potentials start from a position close to stable in the infinite past at the stage of inflationary compression, experience fluctuations near the minimum point of the scale factor and then return to an equilibrium position in the infinite future already at the stage of inflationary expansion.
\FigH{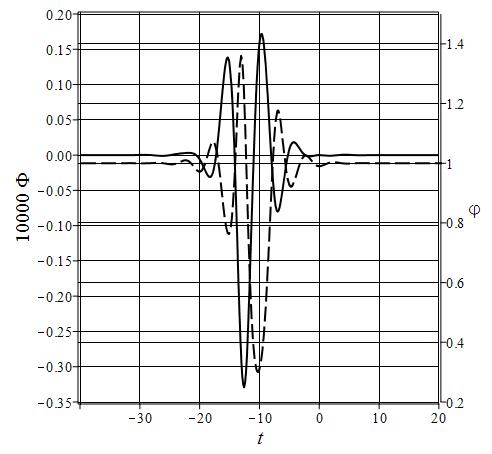}{8}{6.5}{\label{Ign2_6}Scalar potential $\Phi(t)$ -- solid, $\varphi(t)$ -- dashed line for initial conditions $\mathbf{I^+_1}$; $\gamma=10^{-5}$.}
\subsection{Effect of the constant $\gamma$}
Note that if the initial conditions deviate from the coordinates of the points $M_{0,\pm1}^{-}$, the value of the parameter $\gamma$ begins to influence the behavior of the scale functions $\xi(t)$ and $H(t)$. In Fig.\ref{Ign2_4}--\ref{Ign2_5} just such an influence was demonstrated. As can be seen from these graphs, as the parameter $\gamma$ increases, the cosmological model acquires an initial singularity at a finite value of cosmological time (in our case,
Fig.\ref{Ign2_5}, $t_s\approx -8.0421765$). At the same time, the Universe finds its beginning, lost by endless inflation in an unstable state.
\Fig{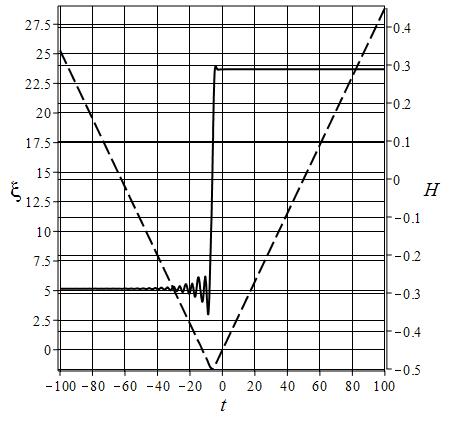}{8.5}{\label{Ign2_4}Function $\xi(t)$ is dashed, $H(t)$ is solid line for initial conditions $\mathbf{I^+_1}$; $\gamma=1$.}
\FigH{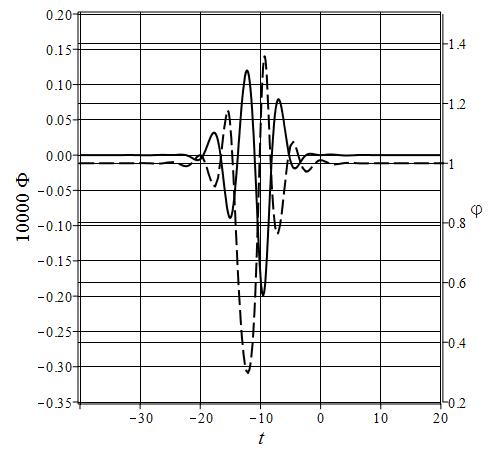}{8}{6.5}{\label{Ign2_7}Scalar potential $\Phi(t)$ -- solid, $\varphi(t)$ -- dashed line for initial conditions $\mathbf{I^+_2}$; $\gamma=10^{-5}$.}

At $\gamma=1$, the scenario with the rebound point remains, but the graphs of $\xi(t)$ and $H(t)$ near the rebound point become more contrasting. At $\gamma=10$ the rebound point turns into a cosmological singularity.

In Fig.\ref{Ign2_8} shows the influence of the value of the constant $\gamma$ on the evolution of scalar potentials. This case, as we answered above (see Fig.\ref{Ign2_4} -- \ref{Ign2_5}), corresponds to a Universe with a rebound point, or with an initial singularity at $t_s\approx-8.0421765$.

The short-term generation of a classical field near the rebound point or singularity occurs precisely due to the kinetic connection of the classical and phantom fields.

\Fig{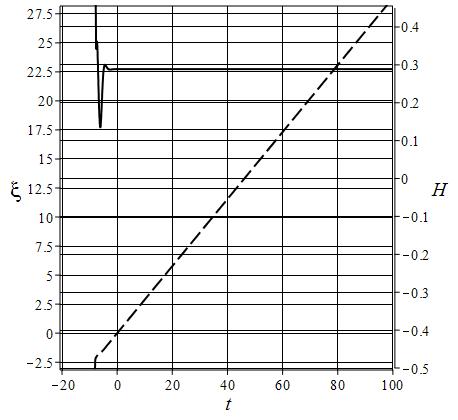}{8.5}{\label{Ign2_5}Function $\xi(t)$ is dashed, $H(t)$ is solid line for initial conditions $\mathbf{I^+_1}$; $\gamma=10$.}
\Fig{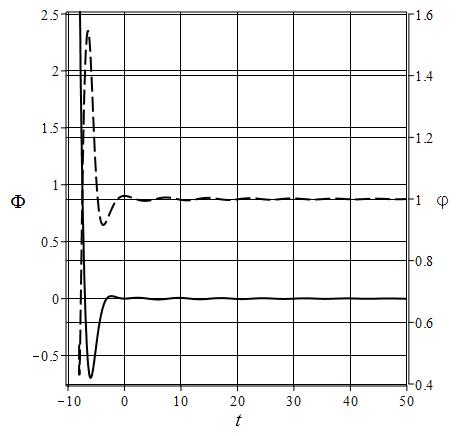}{8.5}{\label{Ign2_8}Scalar potential $\Phi(t)$ -- solid, $\varphi(t)$ -- dashed line for initial conditions $\mathbf{I^+_2}$; $\gamma=1$.}

In Fig.\ref{Ign2_9} shows graphs of the evolution of the barotropic coefficient of cosmological matter $\kappa(t)$ \eqref{kappa} for the initial conditions $\mathbf{I^+_1}$ and $\mathbf{I^+_2}$ at $ \gamma=10^{-5}$. In the first case, the model has a rebound point (Fig.\ref{Ign2_4}), in the second case, an initial singularity at time $t_s\approx -8.0421765$ (Fig.\ref{Ign2_5}). The system passes through the state $H=$, due to which singularities of the functions $\Omega(t)$ and $\kappa(t)$ arise at this point. Far from this critical point, the cosmological model is in the inflationary regime ($\kappa=-1\Rightarrow\Omega=1$). Among other things, you can see that the model approaches the singularity with the barotropic coefficient $\kappa=1$.
\Fig{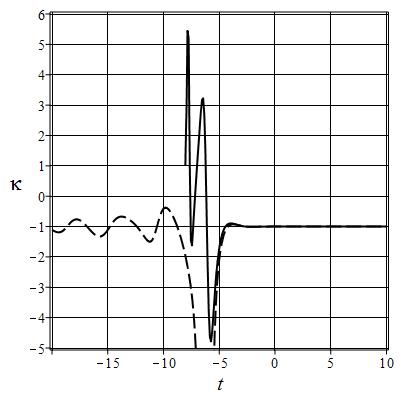}{8.5}{\label{Ign2_9}Functions of the barotropic coefficient $\kappa(t)$ \eqref{kappa}: solid line for the initial conditions $\mathbf{I^+_1}$, dashed line -- $\mathbf{I^+_2}$; $\gamma=1$.}
\section{Model with a finite past}
The models with an initial singularity, i.e., with a finite past, discussed in the previous section are more likely demonstration examples than models of the real Universe. Indeed, the value of the kinetic interaction constant between the components $\gamma\gtrsim 1$ with values of the other fundamental constants in the set $\mathbf{P_1}$ of order 1 \eqref{P_1} is unacceptably large. Indeed, with large-scale transformations of such a model to a real model with masses of superheavy bosons of the order of $10^{17}G$, according to Remark I.\ref{mark2} we obtain for this model
\[\tilde{\mathbf{P}}=\bigl[\bigl[10^{-4},10^{-4},10^{-2},10^{-2}\bigr],1,3\cdot10^{-10}\bigr].\]
Thus, the kinetic interaction constant of the components turns out to be 4 orders of magnitude larger than the self-interaction constants of the classical $\alpha$ and phantom $\beta$ fields, which is physically unacceptable. The real value of the interaction constant must be less than or on the order of the value of the self-interaction constants of these fields.

Therefore, in the class of models with parameters $\mathbf{P_1}$ we consider especially the model with parameters
\begin{equation}\label{P^*_1}
\mathbf{P^*_1}=[[1, 1,1,1],10^{-5},3\cdot10^{-6}]
\end{equation}
and initial conditions
\begin{equation}\label{I_1}
\mathbf{I_1}=[0.9,0,0.9,0,1].
\end{equation}
\subsection{Evolution of geometric factors}
This model with an infinite inflationary future with the Hubble parameter $H_{+\infty}\approx 0.289$ (stable equilibrium point $M_{0,\pm1}^{+}$, Tab.\ref{Tab1}) and a finite past, which corresponds to the singularity at time $t=t_s\approx -2.2284930$. Fig.\ref{Ign2_10} shows the evolution of the scale function $\xi(t)=\ln(a(t))$ and the Hubble parameter $H(t)$ \eqref{sys0a} for this model. %
The coordinates and characters of singular points obviously coincide with the corresponding values indicated in Tab.\ref{Tab1}.

On the graph of the function $H(t)$ Fig.\ref{Ign2_10}, two plateaus can be identified in the intervals $T_1\approx (-1,2)$ and $T_2\approx (5,\infty)$, which exactly correspond to singular points: $T_1$ -- saddle points $M_{\pm1,\pm1}^{+}$, $T_2$ -- attracting points $M_{0,\pm1}^{+}$ and saddle points $M_ {\pm1,0}^{+}$. The $H(t)$ curve demonstrates the transition from an unstable state (saddle point $M_{\pm1,\pm1}^{+}$) through the first plateau to a stable state (attracting point $M_{0,\pm1}^{+ }$) - second plateau.

\Fig{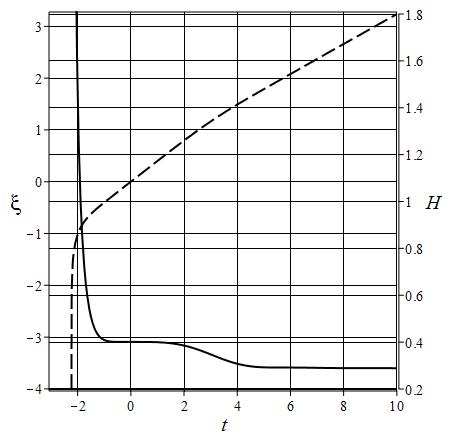}{8}{\label{Ign2_10}Evolution of the scale function $\xi(t)$ (dashed) and the Hubble parameter $H(t)$ (solid) in a model with a finite past at the point $t_s\approx -2.2284930$ with parameters $\mathbf{P^*_1} $ \eqref{P^*_1} and initial conditions $\mathbf{I_1}$ \eqref{I_1}.}

In Fig.\ref{Ign} the solid line shows the behavior of the barotropic coefficient $\kappa(t)$ near this singularity, calculated by the formula \eqref{kappa} using expressions for the total energy density \eqref{E_m} and pressure \eqref{P_m}. Note that the value of this coefficient, calculated both from the indicated formulas and from the formula for cosmological acceleration \eqref{Omega}, coincide. Further in the same figure, dashed and long-dashed lines show the behavior of the barotropic coefficients for the classical and phantom components, respectively; the dotted line $\kappa=1$ shows the behavior of the barotropic coefficient for the interaction component. According to this graph, at the singularity point the barotropic coefficient reaches the value $\kappa=1$, which corresponds to the total extremely rigid equation of state $P=E$ \eqref{kappa}. Over time, the functions $\kappa(t)$ and $H(t)$ tend to $\kappa\to-1$, which corresponds to inflationary compression with the inflationary equation in the state $P=-E$.
\Fig{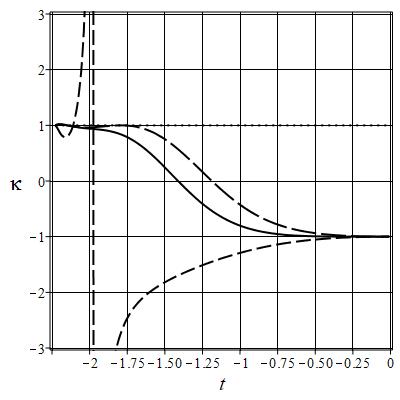}{8.5}{\label{Ign}Evolution of barotropic coefficients $\kappa(t)$ for various system components in a model with parameters $\mathbf{P^*_1}$ \eqref{P^*_1} and initial conditions $\mathbf{I_1}$ \eqref{I_1}.}

\subsection{Evolution of scalar fields}
In the future, we will be faced with the need to display graphs on significantly different scales. In order to overcome this problem, we will use a one-to-one scaling mapping if necessary (see \cite{Yu_Ass}):
\begin{eqnarray}
&\rm{Lig}(x)=\sgn(x)\log[10](1+|x|); \nonumber\\
 &\rm{Lig}^{-1}(x) =\sgn(x)\left(10^{|x|}-1\right).
\end{eqnarray}
\FigH{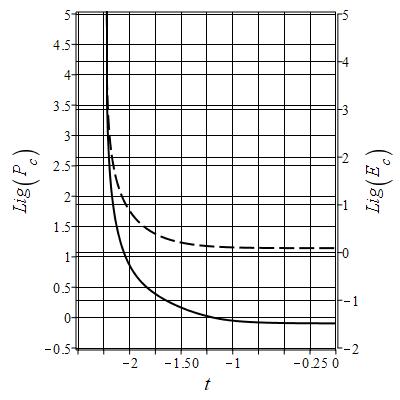}{8}{6}{\label{Ign2_11}Evolution of energy density and pressure.}

Fig.\ref{Ign2_11} shows graphs of the evolution of energy density and pressure of the classical field, according to their contributions to the expressions \eqref{E_m} and \eqref{P_m}: $E_c(t)$ (solid line) and $P_c(t )$ (dashed line) of a classical scalar field in a model with a finite past at the point $t_s\approx -2.2284930$ with parameters $\mathbf{P^*_1}$ \eqref{P^*_1} and initial conditions $\mathbf{ I_1}$ \eqref{I_1}. Note that these quantities are singular at the point of cosmological singularity.

Fig.\ref{Ign2_12} shows graphs of the evolution of the scalar classical $\Phi(t)$ and phantom $\varphi(t)$ potentials, and Fig.\ref{Ign2_13} shows the derivatives of these potentials.

\FigH{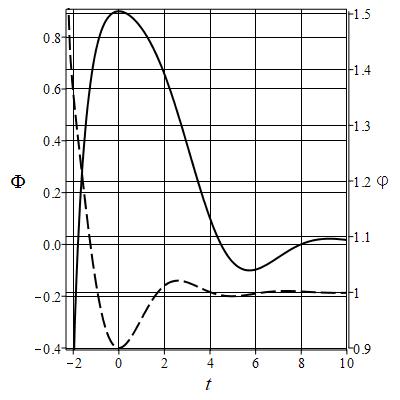}{8}{6}{\label{Ign2_12}Evolution of $\Phi(t)$ (solid line) and $\varphi(t)$ (dashed line) in the model with parameters $\mathbf{P^*_1}$ \eqref{P^*_1} and initial conditions $\mathbf{I_1}$ \eqref{I_1}.}
\FigH{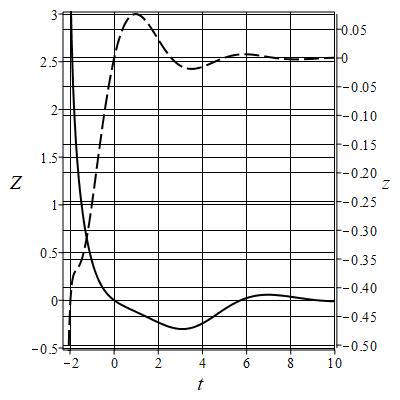}{8}{6}{\label{Ign2_13}Evolution of $Z(t)$ (solid line) and $z(t)$ (dashed line) in a model with parameters $\mathbf{P^*_1}$ \eqref{P^*_1} and initial conditions $\mathbf {I_1}$ \eqref{I_1}.}
\section{A model with a finite future}
Let us now consider a model with parameters
\begin{eqnarray}\label{P_2}
\hspace{-2mm}\mathbf{P_2}=\biggl[[0.01,0.01,0.1,0.1],
0.09876065,0.01\biggr]
\end{eqnarray}
and initial conditions
\begin{equation}\label{I_2}
\mathbf{I_2}=[0,0,1,0.01,-1].
\end{equation}
\subsection{Singular points}
\begin{center}
\refstepcounter{table} {\large Tab. \thetable. \label{Tab2} Character (type) of singular points \linebreak in subspaces $[\Sigma_\Phi,\Sigma_\varphi]$}in the model with parameters $\mathbf{P_2}$\\[12pt]
\begin{tabular}{|c|c|c|}
\hline
Singular points & Coordinates & Type \\
\hline
$M_{0,0}^{+}$ & $[0,0,0,0,0.0577]$ &  $[\mathbf{A},\ \mathbf{S}]$ \\
\hline
$M_{0,0}^{-}$ & $[0,0,0,0,-0.0577]$ & $[\mathbf{R},\ \mathbf{S}]$\\
\hline
$M_{0,\pm1}^{+}$ & $[0,0,\pm1,0,0.0645]$ & $[\mathbf{A},\ \mathbf{A}]$ \\
\hline
$M_{0,\pm1}^{-}$ &  $[0,0,\pm1,0,-0.0645]$ & $[\mathbf{R},\ \mathbf{R}]$ \\
\hline
$M_{\pm1,0}^{+}$ & $[\pm1,0,0,0,2.887]$ & $[\mathbf{S},\ \mathbf{S}]$ \\
\hline
$M_{\pm1,0}^{-}$ & $[\pm1,0,0,0,-2.887]$ &   $[\mathbf{S},\ \mathbf{S}]$ \\
\hline
$M_{\pm1,\pm1}^{+}$ & $[\pm1,0,\pm1,0,0.0707]$ & $[\mathbf{S},\ \mathbf{A}]$ \\
\hline
$M_{\pm1,\pm1}^{-}$ & $[\pm1,0,\pm1,0,-0.0707]$ &  $[\mathbf{S},\ \mathbf{R}]$ \\
\hline
\end{tabular}
\end{center}
\subsection{Evolution of geometric factors}
Fig.\ref{Ign2_14} shows the evolution of the scale function $\xi(t)=\ln(a(t))$ and the Hubble parameter $H(t)$ \eqref{sys0a} for this model.
\FigH{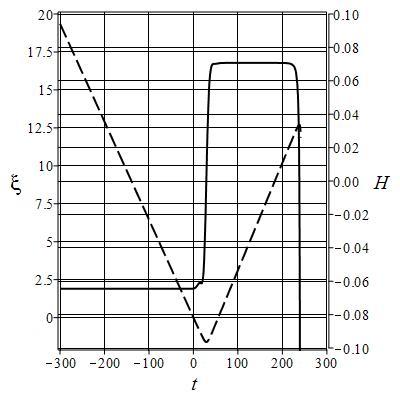}{8}{7}{\label{Ign2_14}Evolution of $\xi(t)$ (solid line) and $H(t)$ (dashed line) in a model with a finite future at point $t=t_s\approx 241$ with parameters $\mathbf{P_2}$ \eqref{P_2} and initial conditions $\mathbf{I_2}$ \eqref{I_2}.}

This model has a rebound point at $t\approx29$, after which it enters the expansion stage $t\in(30,240)$ with positive inflation with $H_0\approx +0.065$, after which it goes into a singular state such as a Big Rip , see, for example, \cite{BigRip}).

\FigH{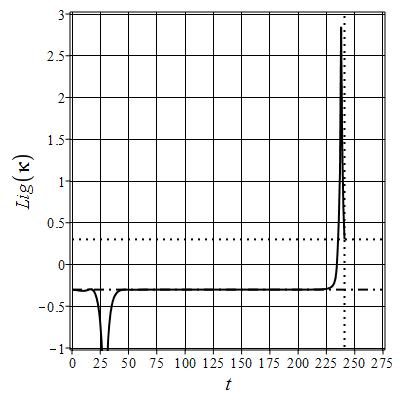}{8}{7}{\label{Ign2_15} Graph of $\kappa(t)$ in a model with a finite future at point $t=t_s\approx 241$ with parameters $\mathbf{P_2}$ \eqref{P_2} and initial conditions $\mathbf{I_2}$ \eqref{I_2}.}

In Fig.\ref{Ign2_15} shows a graph of barotropic coefficients $\kappa(t)$. On this graph, the vertical dotted line corresponds to the singularity $t=t_s\approx 241$ - the break point, the horizontal dotted line corresponds to the value $\kappa=1$, the horizontal dash - the dotted line - to the value $\kappa=-1.$

\FigH{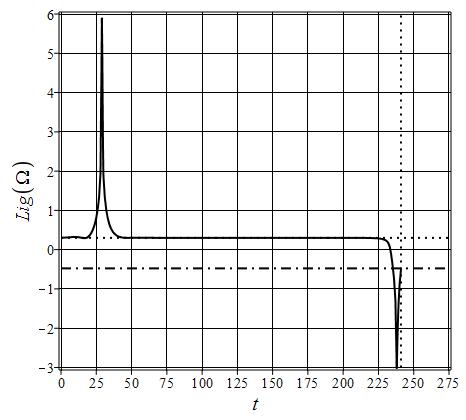}{8}{7}{\label{Ign2_16} Graph of $\Omega(t)$ in a model with a finite future at point $t=t_s\approx 241$ with parameters $\mathbf{P_2}$ \eqref{P_2} and initial conditions $\mathbf{I_2}$ \eqref{I_2}.}

In Fig.\ref{Ign2_16} shows a graph of the invariant cosmological acceleration $\Omega(t)$. On this graph, the vertical dotted line corresponds to the singularity $t=t_s\approx 241$ - the break point, the horizontal dotted line corresponds to the value $\Omega=1$, the horizontal dash - the dotted line - to the value $\Omega=-2$.

Thus, the cosmological model has a singularity in the future corresponding to a large gap, and near the singularity this kind of model behaves in the same way as near the initial singularity: $\kappa\to1$,
$\Omega\to-2$.

\subsection{Evolution of scalar fields:\newline generation of a classical field\newline phantom}
Let us consider the problem of the cosmological evolution of the scalar fields $\Phi(t)$ and $\varphi(t)$. In particular, we are interested in the possibility of generation of one of the components of a scalar doublet by another due to the kinetic connection between them. This question was posed in the first part of the article \cite{Part1}.

Fig.\ref{Ign2_17} shows the graphs
evolution of the scalar classical $\Phi(t)$ and phantom $\varphi(t)$ potentials, and in Fig.\ref{Ign2_18} - the derivatives of these potentials. In the given graphs of these figures one can observe the transition
cosmological model from a state of inflationary compression in the infinite past, which, according to Tab.\ref{Tab2} corresponds to a point $M^-_{0,1}$ of type $\mathbf{[R,R]}$ with coordinates $\Phi(-\infty)=0$, $\varphi(-\infty)=1$, into the state of inflationary expansion, which corresponds to a point $M^+_{1,-1}$ of type $\mathbf{[S,A]}$ with coordinates $\Phi=1$, $\varphi =-1$. This state ends with a Big Rip.

\Fig{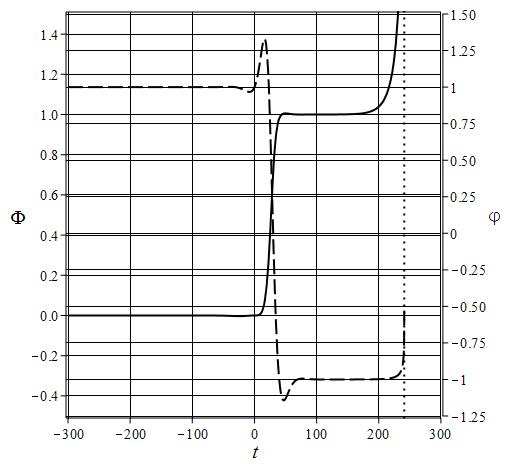}{9}{\label{Ign2_17}Evolution of scalar classical $\Phi(t)$ (solid) and phantom $\varphi(t)$ (dashed) potentials with parameters $\mathbf{P_2}$ \eqref{P_2} and initial conditions $\mathbf{I_2}$ \eqref{I_2}.}
\Fig{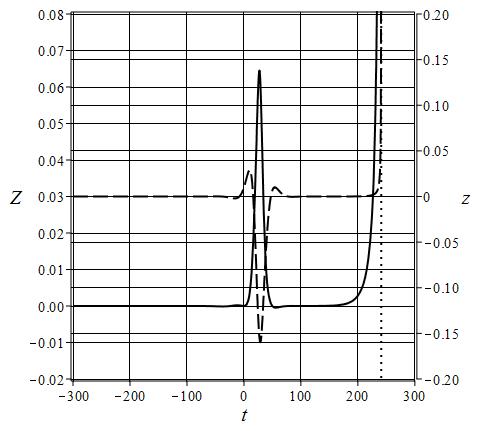}{9}{\label{Ign2_18}Evolution of dynamic functions $Z(t)$ (solid) and $z(t)$ (dashed) in the model with parameters $\mathbf{P_2}$ \eqref{P_2} and initial conditions $\mathbf{I_2}$ \eqref{I_2}.}

Thus, in this model, at the stage of inflationary expansion before the Big Rip, the generation of a classical scalar field is observed. This process corresponds to a transition with a stable state for the phantom field and an unstable state for the classical one.

\section{A model with a finite past and future}
Let's consider a model with parameters $\mathbf{P_2}$ \eqref{P_2} and initial conditions very close to the coordinates of the singular point $M^+_{1,0}$, according to Tab.\ref{Tab2} saddle point of type $\mathbf{[S,\ S]}$, namely, just above it (only $10^{9}$ !)
\begin{equation}\label{I_3}
\mathbf{I_3}=[1+10^{-9},0,0,0,1].
\end{equation}
This is a model with a finite past, which corresponds to a singularity at the time $t=t^0_s\approx -84.206597$, and a finite future, which corresponds to a singularity at the time $t=t^1_s\approx 280.04421$.
Fig.\ref{Ign2_19} shows the evolution of the scale function $\xi(t)=\ln(a(t))$ and the Hubble parameter \eqref{sys0a} $H(t)$ for this model.
The coordinates of singular points and their characters are indicated in Tab.\ref{Tab2}. Thus, the graphs in Fig.\ref{Ign2_19} demonstrate the transition from an unstable state (saddle point $M_{0,0}^{+}$) through an inflationary plateau to the final singular state (Big Rip).

\FigH{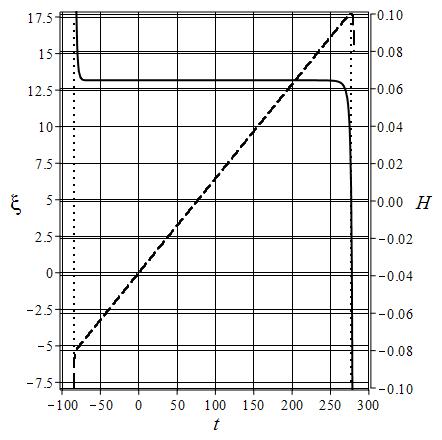}{8}{7}{\label{Ign2_19}Evolution of the scale function $\xi(t)$ is dashed and the Hubble parameter $H(t)$ is solid line in the model with a finite past $t=t^0_s\approx -84.206597$ (left vertical dotted line) and a finite future $t=t^1_s\approx 280.04421$ (right vertical dotted line) with parameters $\mathbf{P_2}$ \eqref{P_2} and initial conditions $\mathbf{I_3}$ \eqref{I_3}.}

In Fig. \ref{Ign2_20} shows the behavior of the barotropic coefficient $\kappa(t)$ in this model; in the same figure one can see that the cosmological model at both singular points, $t^0_s$ and $t^1_s$, manifests itself as matter with an extremely rigid equation of state.

\FigH{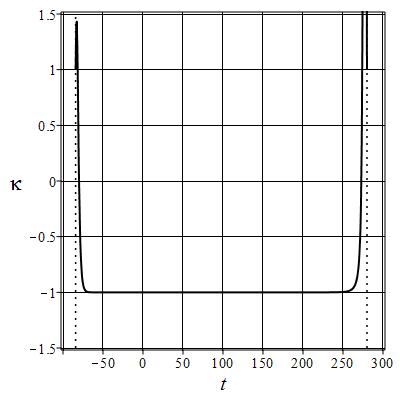}{8}{7}{\label{Ign2_20}Evolution of the barotropic coefficient $\kappa(t)$ in a model with parameters $\mathbf{P_2}$ \eqref{P_2} and initial conditions $\mathbf{I_3}$ \eqref{I_3}.}

In Fig.\ref{Ign2_21}--\ref{Ign2_22} the behavior of the barotropic coefficient near singular points is shown on a large scale. %
Finally, Fig.\ref{Ign2_23} shows graphs of the evolution density of the evolution of the potentials of the classical field and phantom fields for this model. As we noted in \cite{Part1}, at the cosmological singularity points $t^0_s$ and $t^1_s$ these potentials are also singular. Outside these points on the inflation interval $t\in(t^0_s,t^1_s)$ the potential values are close to the coordinates of the saddle point $M^+_{0,0}$.
\FigH{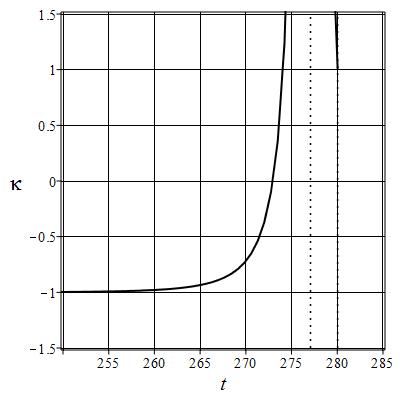}{8}{7}{\label{Ign2_21}Evolution of barotropic coefficients $\kappa(t)$ in a model with parameters $\mathbf{P_2}$ \eqref{P_2} and initial conditions $\mathbf{I_3}$ \eqref{I_3}.}
\FigH{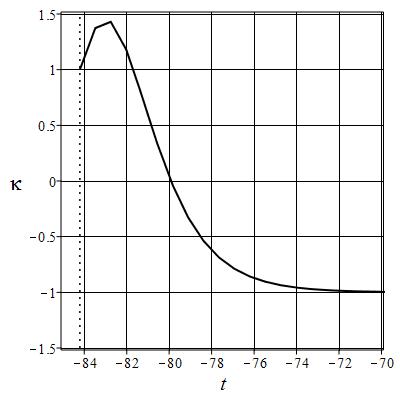}{8}{7}{\label{Ign2_22}Evolution of barotropic coefficients $\kappa(t)$ in a model with parameters $\mathbf{P_2}$ \eqref{P_2} and initial conditions $\mathbf{I_3}$ \eqref{I_3}.}

Note that the lifetime of this cosmological model $\Delta t=t^1_s-t^0_s$ is sensitive to the difference between the initial value of the potential of the classical scalar field and its value at the unstable point $M_{+1,0}^{+}$ -- $\Delta\Phi_0\equiv\Phi_0-1$. As $\Delta\Phi_0$ decreases, the model lifetime increases according to the empirical law
\[\Delta t\approx 41\cdot|\lg(\Delta\Phi_0)|.\]
However, the main resource for increasing the lifespan of a model lies in its large-scale transformations. By choosing the similarity coefficient $k=10^2$, we will achieve an increase in the model lifetime by exactly $k$ times.
\FigH{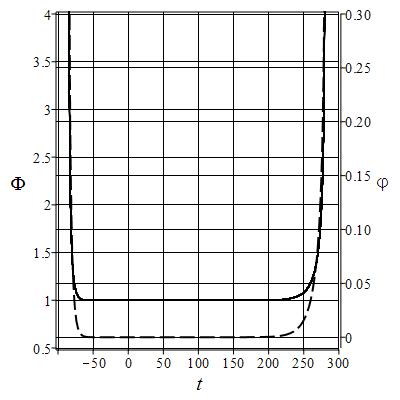}{8}{7}{\label{Ign2_23}Evolution of potentials $\Phi(t)$ is solid and $\varphi(t)$ is dashed in the model with parameters $\mathbf{P_2}$ \eqref{P_2} and initial conditions $\mathbf{I_3}$ \eqref{I_3}.}

\section{Generating a classical scalar field near the rebound point}
As the results of numerical modeling show, a stable classical scalar field corresponding to an inflationary expansion appears to be generated only in models with a finite past at the stage between the rebound point and the Big Rip. In other cosmological models, the classical scalar field is generated only at intermediate stages of cosmological evolution near the rebound point, if there is one. The existence of a rebound point implies that before this point the model was in the compression stage, and after the rebound point it switched to the expansion stage. After this, the system goes into an equilibrium state, the classical scalar field disappears, and further inflation is supported only by the phantom field in stable equilibrium. Let us consider this process in more detail using a specific example of a model with an infinite past and an infinite future:
\begin{eqnarray}\label{P_3}
\mathbf{P_3}=\bigl[[0.01,0.01,0.1,0.1],10^{-2},10^{-6}\bigr];\\
\label{I_4}
\mathbf{I_4}=[0,0,0.01,0,1].
\end{eqnarray}
Thus, at time $t=0$ there is no classical scalar field, and the phantom one is very small.

\subsection{Singular points and geometric factors}
The singular points of the model are shown in Tab.\ref{Tab3}.
\begin{center}
\refstepcounter{table} {\large Tab. \thetable. \label{Tab3} Characters (type) of singular points\linebreak in subspaces $[\Sigma_\Phi,\Sigma_\varphi]$} in the model with parameters  $\mathbf{P_3}$\\[12pt]
\begin{tabular}{|c|c|c|}
\hline
Singular points & Coordinates & Type \\
\hline
$M_{0,0}^{+}$ & $[0,0,0,0,0.000576]$ &  $[\mathbf{A},\ \mathbf{S}]$ \\
\hline
$M_{0,0}^{-}$ & $[0,0,0,0,-0.000576]$ & $[\mathbf{R},\ \mathbf{S}]$\\
\hline
$M_{0,\pm1}^{+}$ & $[0,0,\pm1,0,0.0289]$ & $[\mathbf{A},\ \mathbf{A}]$ \\
\hline
$M_{0,\pm1}^{-}$ &  $[0,0,\pm1,0,-0.0289]$ & $[\mathbf{R},\ \mathbf{R}]$ \\
\hline
$M_{\pm1,0}^{+}$ & $[\pm1,0,0,0,2.89]$ & $[\mathbf{S},\ \mathbf{S}]$ \\
\hline
$M_{\pm1,0}^{-}$ & $[\pm1,0,0,0,-2.89]$ &   $[\mathbf{S},\ \mathbf{S}]$ \\
\hline
$M_{\pm1,\pm1}^{+}$ & $[\pm1,0,\pm1,0,0.0409]$ & $[\mathbf{S},\ \mathbf{A}]$ \\
\hline
$M_{\pm1,\pm1}^{-}$ & $[\pm1,0,\pm1,0,-0.0409]$ &  $[\mathbf{S},\ \mathbf{R}]$ \\
\hline
\end{tabular}
\end{center}

In Fig. \ref{Ign2_24} shows graphs of the evolution of the geometric factors $\xi(t)$ and $H(t)$.
\FigH{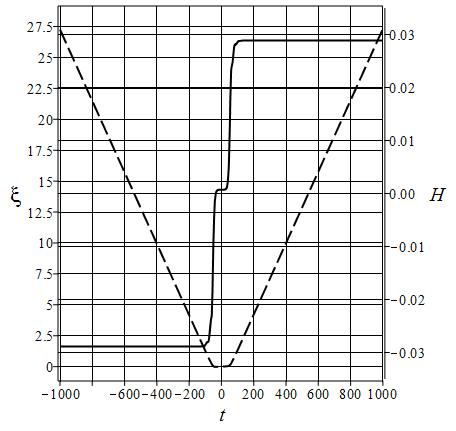}{8}{7}{\label{Ign2_24}Evolution of geometric factors in a model with parameters $\mathbf{P_3}$ and initial conditions $\mathbf{I_4}$: $\xi(t)$ -- dashed and $H(t)$ -- solid line.}
\subsection{Evolution of scalar fields}
In Fig.\ref{Ign2_25} shows graphs of the cosmological evolution of the components of the scalar doublet in this model.
\FigH{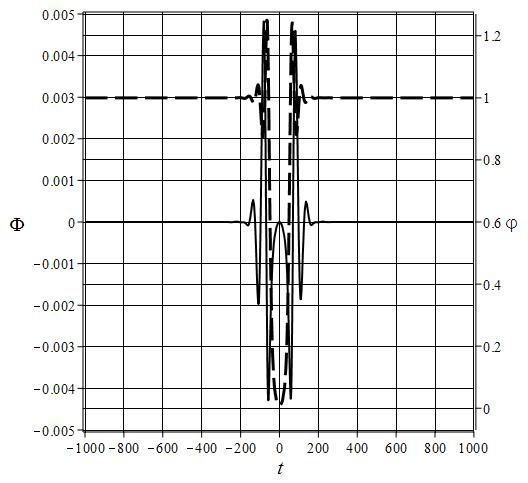}{8}{7}{\label{Ign2_25}Evolution of scalar fields in a model with parameters $\mathbf{P_3}$ and initial conditions $\mathbf{I_4}$: $\varphi(t)$ is dashed and $\Phi(t)$ is solid line.}

Thus, in the infinite past, the Universe starts from a state \emph{close} to the singular point $M^-_{0,1}=[0,0,1,0,-0.0289]$ and is in a state of inflationary compression. Near the rebound point $t_b\approx0$
oscillations of the phantom and classical scalar fields arise in the system, after which the system goes into a state of inflationary expansion corresponding to the singular point $M^+_{0,1}=[0,0,1,0,+0.0289]$, restoring the original values of potentials of scalar fields.

In Fig.\ref{Ign2_26}--\ref{Ign2_27} the process of oscillation of scalar field potentials near the rebound point is shown in close-up.
\FigH{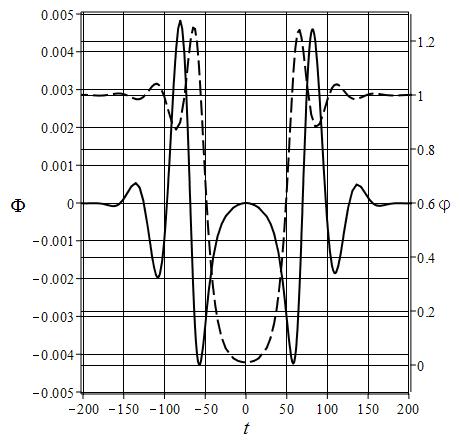}{8}{7}{\label{Ign2_26}Evolution of scalar fields in a model with parameters $\mathbf{P_3}$ and initial conditions $\mathbf{I_4}$: $\Phi(t)$ -- solid and $\varphi(t)$ -- dashed line (close-up).}

\FigH{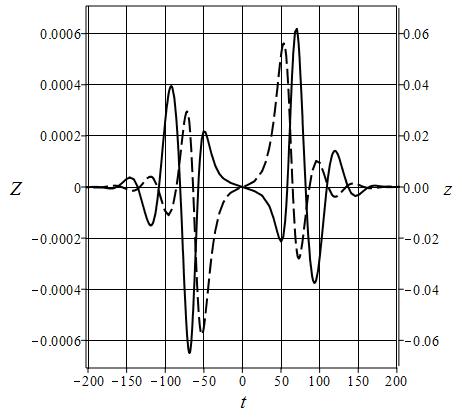}{8}{7}{\label{Ign2_27}Evolution of derivatives of scalar fields in a model with parameters $\mathbf{P_3}$ and initial conditions $\mathbf{I_4}$: $Z(t)\equiv\dot{\Phi}$ -- solid and $z(t)\equiv\dot{\varphi}$ -- dashed line (close-up)}

In this model, the interval of oscillation of scalar field potentials, at which the potential of the classical field reaches values of the order of $|\Phi(t_b)|\sim 0.005$, and the potential of the phantom field becomes zero, takes about 300 Planck times. When using a scaling transformation to real values of fundamental constants for the scale of QCD theory, this period can stretch to $3\cdot10^4$ Planck times, and for the Standard Model - up to $3\cdot10^{15}t_{Pl}$.
\subsection{Production of scalarly charged\newline fermions near the rebound point}
In the above graphs of the evolution of scalar fields (Fig. \ref{Ign2_6}, \ref{Ign2_7}, \ref{Ign2_12}, \ref{Ign2_13}, \ref{Ign2_16}, \ref{Ign2_17}, \ref{Ign2_25} and \ref{Ign2_26}) sufficiently large and fast bursts of the potential of the classical scalar field $\Phi$ and its derivative $z(t)$ are observed near the rebound points (or singularity points, if they exist). If we do not take into account exotic cosmological models with a finite future, then the models presented in the above figures reveal in the region of the rebound points the amplitude of oscillations of the derivative of the potential $\max|\dot{\Phi}|\sim 2\cdot10^{-6}\div 6\cdot10^{-4}$.

Let us estimate the rate of production, for example, of pairs of scalarly charged fermions with charge $q$ and mass $m_f$ in a classical scalar field with a time derivative $Z_0$, using an analogy with the formula for the rate of creation of an electron-positron pair in an electric field of strength $E$ (see, for example, \cite{Shwinger})
\begin{eqnarray}\label{w_ee}
\dot{n}_{ee_+}=\frac{e^2E^2}{4\pi^3\hbar^2c^2}\displaystyle\mathrm{e}^{-\frac{E_0}{E}},\nonumber
\end{eqnarray}
where $w_{ee_+}$ is the rate of electron-positron pair production per unit volume, $m$ is the mass of the electron, $e$ is its charge, $E_c$ is the critical value of the electric field strength
\begin{eqnarray}
E_0=\frac{\pi m^2c^3}{\hbar e}.\nonumber
\end{eqnarray}
Passing to the Planck units in these formulas and making the substitutions $e\to q$, $m\to m_f$ and $E\to Z$, we obtain in the adiabatic approximation an estimate of the rate of production of pairs of scalarly charged fermions
with scalar charge $q$ and mass $m_f$ in scalar field $\Phi(t)$
\begin{eqnarray}\label{w_ff}
\dot{n}_{qq_+}(t)\backsimeq \frac{q^2Z^2(t)}{4\pi^3}\displaystyle\exp\left(-\frac{\pi m^2_f}{|qZ(t)|}\right).
\end{eqnarray}

For efficient production of fermion pairs, the following condition must be met:
\begin{eqnarray}\label{Z>Z_0}
|Z(t)|\gtrsim \frac{\pi m^2_f}{|q|},
\end{eqnarray}
that is, fermions must be light, but at the same time have a sufficiently large scalar charge.
The final density of generated scalarly charged fermions is obtained by integrating over time the expression \eqref{w_ff}:
\begin{eqnarray}\label{n_qq}
n_{qq_+}=\int\limits_{t_0}^{t_0+\Delta t}\frac{q^2Z^2(t)}{4\pi^3}\displaystyle\exp\left(-\frac{\pi m^2_f}{|qZ(t)|}\right)dt  \nonumber\\
\backsimeq \frac{q^2\overline{Z^2(t)}}{4\pi^3}\displaystyle\exp\left(-\frac{\pi m^2_f}{|q\overline{Z(t)}|}\right)\Delta t,
\end{eqnarray}
where $\Delta t$ is the duration of the burst of the classical field potential, $\overline{Z(t)}$ is the root-mean-square value of the function $Z(t)$ on the burst interval.

Let us evaluate the possibility of the creation of fermion pairs based on the last model we studied $\mathbf{P_3}$ \eqref{P_3}, subjecting it, according to Property I.2, to scale transformations in order to bring the model parameters closer to real parameters, for example, $\textbf{SU(5)}$. At the same time, we must also transform the value of the scalar charge according to \cite{TMF_24} according to the law $\tilde{q}=q/\sqrt{k}$. Thus, after a scaling transformation with a similarity coefficient $k$, the expression \eqref{n_qq} is transformed to the form:
\begin{eqnarray}\label{kw_ff}
n_{qq_+}(t,k)\backsimeq \frac{q^2\overline{Z^2(t)}}{4\pi^3k^2}\displaystyle\exp\left(-\frac{\pi m^2_f}{\sqrt{k}|q|\overline{|Z(t|)}}\right)\Delta t,
\end{eqnarray}
where $\overline{|Z(t)|}$ is the average value of the modulus of the burst of the derivative of the scalar field $\Phi(t)$.

The function $n_{qq_+}(t,k)$ has a maximum at
\begin{eqnarray}\label{k_max}
k=\left(\frac{\pi m^2_f}{4|q|\overline{|Z(t)|}}\right)^2,
\end{eqnarray}
from where we get the following:
\begin{eqnarray}\label{n_max}
\max(n_{qq_+}(t))\backsimeq \frac{q^4\overline{Z^2(t)} \mathrm{e}^{-4}}{4\pi^5m^4_f}.
\end{eqnarray}
At $q=10^{-2}$, $m_f=10^{-4}$ and according to Fig.\ref{Ign2_27}  $\overline{|Z(t)|}\backsimeq 2\cdot 10^{-7}$ we obtain from \eqref{n_max} \linebreak $\max(n_{qq_+}(t))\backsimeq 3\cdot 10^{-10}$. At first glance, this value seems very small, but when assessing it, two important circumstances must be taken into account. Firstly, this value is expressed in Planck units of length; in Compton units of length $\lambda_f$ relative to the generated fermions this value will already be $300\lambda^{-3}_f$). Secondly, it is necessary to take into account the factor of increase in volume as a result of cosmological expansion $\exp(-3\xi(t))$, which, according to the graph in Fig.\ref{Ign2_24} is about
$\exp(-3\cdot3)\thickapprox 10^{-4}$. As a result, upon completion of the process of fermion pair production, we obtain their concentration of the order of $n_{ff_+}\gtrsim 3\cdot 10^{-2}\lambda^3_f$. In addition, the statistical factor $N_f$ of the number of fermion types in the interaction model is also important. In the standard SU(5) model (colors, electric charges) $N_f\sim 10$, which increases the estimate of the total number of fermion pairs by another order of magnitude.

\section{Conclusion}
Summing up the main results of both parts of the work, we indicate its following main results.\\[12pt]
$\bullet$ A model of the evolution of the Universe based on an asymmetric Higgs scalar doublet with a kinetic coupling proportional to the product of the derivatives of the components of the scalar doublet is proposed and studied.
In particular, a qualitative analysis of the dynamic system of the model was carried out, the symmetry properties of the model with respect to the reflection and similarity transformation were investigated and proven.\\
$\bullet$ The main types of behavior of the cosmological model are revealed depending on the fundamental parameters and initial conditions. Cases of the presence of initial and final cosmological singularities and rebound points have been identified.\\
$\bullet$ It is shown that near the rebound points the phantom component $\varphi$ of a scalar doublet generates its classical component $\Phi$.\\
$\bullet$ The asymptotic behavior of the cosmological model near the points of initial singularity and the Big Rip, corresponding to the barotropic coefficient $\kappa=1$ ($p=\varepsilon$), has been proven analytically and confirmed by numerical modeling.\\
$\bullet$ The probability of fermion pair production in a strong alternating scalar field near the rebound points has been estimated.\\

All identified features of cosmological models based on an asymmetric scalar doublet: the presence of initial singularities, rebound points and large discontinuity points, generation of scalar doublet components, etc., are similar to the features of cosmological models based on systems of scalarly charged fermions \cite{TMF_21} -- \cite{Yu_Ass}, which apparently makes it possible to replace the mathematically complicated model of the formation of supermassive black holes \cite{Ignatev_SBH} -- \cite{TMF_23} with a simpler one, built on a purely field basis. In addition, an important circumstance is also the fairly intense production of fermion pairs by the scalar field
near the rebound points. Thanks to this factor, the required number of scalarly charged fermions can be obtained independently of the processes of gravitational pair production, and at the stage of the cold Universe.

\subsection*{Founding}
The work is performed according to the Russian Government Program of Competitive Growth of Kazan Federal University
%
%%%%%%%%%%%%%%%%%%%%%%%%%%%%%%%%%%%%%%%%%%%%%%%%%

\end{document}